\documentclass[12pt,nohyper,notoc]{JHEP}

\usepackage{amsmath,array,amssymb,cite} % use drftcite in draftmode otherwise
\newcommand{\startappendix}{
\setcounter{section}{0}
\renewcommand{\thesection}{\Alph{section}}}

\newcommand{\Appendix}[1]{
\refstepcounter{section}
\begin{flushleft}
{\large\bf Appendix \thesection: #1}
\end{flushleft}}

\newcommand{\bea}{\begin{eqnarray}}
\newcommand{\eea}{\end{eqnarray}}

\newcommand{\modx}{\ensuremath{|x|}}

\newcommand{\Rowspace}{\phantom{$\Big($}}
\newcommand{\Rsm}{\phantom{\Big(}}

\def\N{{\cal N}}

\def\tr{{\rm tr}}

\def\sst{\scriptscriptstyle}

\def\SU{\text{SU}}

\def\SO{\text{SO}}

\def\USp{\text{USp}}

\def\U{\text{U}}
\def\G{\text{G}}
\def\F{\text{F}}
\def\E{\text{E}}
\def\D{{\cal D}}

\def\Dbarslash{\,\,{\raise.15ex\hbox{/}\mkern-12mu {\bar\D}}}
\def\Dslash{\,\,{\raise.15ex\hbox{/}\mkern-12mu \D}}
\def\delslash{\,\,{\raise.15ex\hbox{/}\mkern-9mu \partial}}
\def\delbarslash{\,\,{\raise.15ex\hbox{/}\mkern-9mu {\bar\partial}}}

\def\VEV#1{\left\langle #1\right\rangle}
\def\Vev#1{\big\langle{#1}\big\rangle}
\let\vev=\Vev

\newcommand{\EQ}[1]{\begin{equation} #1 \end{equation}}
\newcommand{\AL}[1]{\begin{subequations}\begin{align} #1 \end{align}\end{subequations}}
\newcommand{\SP}[1]{\begin{equation}\begin{split} #1 \end{split}\end{equation}}

\renewcommand{\descriptionlabel}[1]%
      {\hspace{\labelsep}\textsf{#1}}
\title{Monopoles, affine algebras and the gluino condensate}

\author{N.~Michael Davies$^{a}$, Timothy J.~Hollowood$^{b}$
and Valentin V.~Khoze$^a$\\
$^a$Department of Physics, University of Durham,
Durham, DH1 3LE, UK\\
$^b$Department of Physics, University of Wales Swansea,
Swansea, SA2 8PP, UK\\
E-mail: {\tt n.m.davies@durham.ac.uk}, {\tt t.hollowood@swan.ac.uk},
{\tt valya.khoze@durham.ac.uk}}

\abstract{We examine the low-energy dynamics of four-dimensional supersymmetric
gauge theories and calculate the values of the gluino condensate for all simple
gauge groups. By initially compactifying the theory on a cylinder we are able to
perform calculations in a controlled weakly-coupled way for small radius. The
dominant contributions to the path integral on the cylinder arise from magnetic
monopoles which play the role of instanton constituents. We find that the
semi-classically generated superpotential of the theory is the affine Toda
potential for an associated twisted affine algebra. We determine the
supersymmetric vacua and calculate the values of the gluino condensate. The
number of supersymmetric vacua is equal to $c_2$, the dual Coxeter number, and
in each vacuum the monopoles carry a fraction $1/c_2$ of topological charge.
As the results are independent of the radius of the circle, they are also valid
in the strong coupling regime where the theory becomes decompactified. In this
way we obtain values for the gluino condensate which for the classical gauge
groups agree with previously known ``weak coupling instanton''  expressions (but
not with the ``strong coupling instanton'' calculations). This detailed
agreement provides further evidence in favour of the recently advocated
resolution of the the gluino condensate puzzle. We also make explicit
predictions for the gluino condensate for the exceptional groups.}


\preprint{{\tt hep-th/0006011}}

\begin{document}

\section{Introduction and summary of results}

The goal of this paper is to provide new calculations of the values of the
gluino condensate $\vev{ \tr \lambda^2}$
in four-dimensional ${\cal N}=1$ supersymmetric Yang-Mills theory
for all the simple gauge groups.
Our results are summarized in the Table 1 and a universal formula in
terms of Lie algebra data is given in \eqref{gcond}. For classical gauge groups
our results are in precise agreement with
the known expressions derived in the ``weak-coupling instanton'' approach in
Refs.~\cite{NSVZtwo,Shifman:1988ia,Morozov,FP}.
For the exceptional gauge groups the condensates are,
to the best of our knowledge, calculated for the first time.
\begin{table}%\setlength{\extrarowheight}{5pt}
\begin{center}\begin{tabular}{ccccc} \hline\hline
\phantom{$\Biggr($} gauge group & $\Lambda^{-3}\vev{ \frac{\tr\lambda^2}{16\pi^2}}$ \\
 \hline
\Rowspace $\SU(N)$ & $1$ \\
\Rowspace $\SO(N)$ &  $2^{\frac{4}{N-2}-1}$ \\
\Rowspace $\USp(2N)$ & $ 2^{1-\frac{2}{N+1}}$  \\
\Rowspace $\G_2$ &  $2^{-\frac{1}{2}}3^{\frac{1}{4}}$\\
\Rowspace $\F_4$ & $2^{-\frac{1}{9}}3^{-\frac{1}{3}}$ \\
\Rowspace $\E_6$ & $2^{-\frac{1}{2}}3^{-\frac{1}{4}}$ \\
\Rowspace $\E_7$ &  $2^{-\frac{7}{9}} 3^{-\frac{1}{3}}$  \\
\Rowspace $\E_8$ & $\qquad2^{-\frac{13}{15}}3^{-\frac{2}{5}} 5^{-\frac{1}{6}}\qquad$\\
\hline\hline
\end{tabular}\end{center}
\caption{\small The values of the gluino condensate in the Pauli-Villars
scheme \eqref{lexct}.}
\end{table}

It is somewhat of a miracle that some features of gauge theories
which have supersymmetry can be understood exactly. Sometimes this
success arises from viewing these theories as being embedded in string
theory, a classical example being the duality of Maldacena \cite{Aharony:2000ti}.
Generally, however, we can make
use of the fact that supersymmetry leads to very restrictive Ward
identities, giving rise to powerful holomorphy properties (see the review \cite{Intriligator:1996au}). Regarding
this later point, the full functional form of certain
correlators is fixed up to an overall constant. Sometimes these
correlators have a
dependence on the couplings which can be identified with specific
gauge theory configurations, in particular with instantons,
but in other cases this is not so \cite{DK}. 
In the
former case, it is tempting to suppose that a semi-classical
instanton calculation will yield the value of the correlator. In
particular,
we have in mind multi-point functions of the gluino
operator $\tr\,\lambda^2$ in $\N=1$ supersymmetrc gauge theory. 
It is our thesis that one must be very
careful in applying a semi-classical analysis to a strongly-coupled
theory and such calculations will only be correct if they are performed
in a weakly-coupled phase, where semi-classical methods are rigorously
justified. It is then possible to infer the value of the correlator in
a strongly-coupled phase, if that phase is continuously connected to
the weakly-coupled phase by using holomorphicity. It was
the misuse of a semi-classical analysis directly in a strongly-coupled phase that
led to the gluino condensate puzzle.

This famous puzzle is the inconsistency between 
two conceptually different approaches followed in the early
literature of calculating the gluino condensate in pure $\N=1$
supersymmetric gauge theory.
In the first methodology \cite{NSVZone,ARV,Amati}---and in the present context 
the ``suspect'' method, because it involves a semi-classical analysis
directly in the strongly-coupled confining phase of the gauge theory---the so-called
strong-coupling instanton (SCI) approach, the gluino condensate
$\Vev{\tr \lambda^2}$ 
is determined via an explicit one-instanton calculation of a certain
multi-point function of $\tr\,\lambda^2$.
Cluster decomposition arguments are then invoked in order to
extract the one-point function $\Vev{\tr \lambda^2}$.
In the second methodology \cite{NSVZtwo},
the so-called weak-coupling instanton (WCI) approach---and for us the
safe method---the calculation is
performed with additional matter fields whose presence ensures that the
non-abelian gauge group is broken and the theory
is in a weakly-coupled Higgs phase and a ``constrained instanton'' calculation
is justified \cite{ADS}.
Holomorphicity is then used to decouple
the matter fields and to flow continuously to the confining phase of
the original gauge theory. As is well known, these two methods give two different
values for the gluino condensate
\cite{NSVZtwo,FS,Amati,FP}:\footnote{These 
results are quoted for an $\SU(N)$ gauge theory
in the Pauli-Villars scheme
with $\Lambda$ being the corresponding dimensional transmutation scale
of the theory defined in Eq.~\eqref{lexct} below.}
\EQ{
\VEV{\tr \lambda^2\over16\pi^2}_{\SU(N)}\ =\
\begin{cases}{2 \over [(N-1)! \ (3N-1)]^{1/N}} \Lambda^3 &
\text{SCI}\ , \\
\Lambda^3 & \text{WCI}\ .
\end{cases}
\label{gcsun}
}
The reason for the discrepancy between the SCI and WCI
calculations, as well as the question as to which is correct,
has been a long-standing controversy
\cite{NSVZtwo,Amati,KS,SVrev}. This controversy was re-examined
in \cite{HKLM} using recently developed multi-instanton
methods \cite{MO,DHKMV}. By evaluating the
$k$-instanton contribution to gluino correlation functions in the large number
of colours limit it was shown that an essential step in the SCI
calculation of the gluino condensate, namely the use of cluster decomposition
in the instanton sector, is actually invalid. The central 
idea pursued in \cite{DHKM} and in the present paper
is that there are additional configurations
which contribute to the gluino condensate 
implying that the SCI calculation only gives part of the answer.
The existence of other contributions to multi-point correlators
of $\tr\,\lambda^2$, which are non-instantonic invalidates
the application of cluster decomposition to a purely instantonic contribution.

In Ref.~\cite{DHKM} we provided an alternative way to deform the theory
in order to connect the confining phase continuously to a weak-coupled
phase: in this case a Coulomb rather than a Higgs phase. 
The idea is to consider the theory partially compactified
on the cylinder ${\mathbb R}^3 \times S^1$. In this scenario, the gauge
field can have a non-trivial Wilson loop around the circle which acts
like an adjoint-valued Higgs field breaking the gauge group
to its maximal abelian subgroup and so the theory is in a
Coulomb phase. For small enough radius,
the resulting theory is arbitrarily weakly coupled
and the gluino condensate can be reliably calculated. It is then
argued, based on the usual argument of holomorphy, 
that the result is actually independent of the radius and is therefore
easily extrapolated to the confining phase where the radius
goes to infinity and the theory becomes decompactified.

However, there is a significant bonus in this scenario: the additional
configurations missing in the instanton calculation can explicitly be
identified at small radius with BPS
monopoles in the gauge theory with the
component of the gauge field around the circle playing the r\^ole of a
Higgs field. However, we should point out that this in no way means
that BPS monopoles quantitatively describe the physics in the
decompactification limit. In this scenario, the one-point function
$\vev{\tr\,\lambda^2}$  directly receives a semi-classical contribution
from single monopoles,
unlike the SCI situation in ${\mathbb R}^4$, where, as described
above, only multi-functions receive contributions. The monopoles consequently
carry fractional topological charge. Hence, the theory on the cylinder
uncovers a very pleasing picture of instanton constituents, or
instanton partons, that were argued to play an important r\^ole in
confinement of ordinary QCD
\cite{Belavin:1979fb,Fateev:1979dc,Berg:1979uq,Callan:1978gz,Osborn}.
The fact that an instanton configuration on the cylinder is 
actually a composite of fundamental monopoles has been the subject of
number of earlier works \cite{LY,KL,Nahmtwo,Garland,LL,KvB}. These generalize
the notion of a periodic instanton, or ``caloron''
\cite{HS,Rossi:1979qe,GPY}, to the case when the gauge field has a
non-trivial Wilson line around the circle. It is only in this case
that the instanton constituents can be pulled apart and identified
with monopoles. It turns out that in ordinary QCD the Wilson line of
the gauge field around the circle is energetically favoured to vanish:
in this case the monopoles have no r\^ole to play in the physics. On
the contrary, as we shall explicitly show, in $\N=1$ supersymmetric
gauge theories, a non-trivial superpotential is generated by the
monopoles whose supersymmetric vacua have a non-trivial value for the
Wilson line and hence monopole effects are important. 
Other recent references which consider supersymmetric
gauge theories  on a cylinder and monopole effects are to be found in 
\cite{SWthree,Aharony:1997bx,Dorey}.

In $\N=1$ supersymmetric gauge theories, the first coefficient of the $\beta$-function is
$b_0=3c_2$, where $c_2$ is the dual Coxeter number of the gauge group
listed in Table 3. We will use a definition of the dynamical scale
$\Lambda$ in the Pauli-Villars renormalization scheme via
the RG-invariant exact relation\footnote{If one chooses to use
instead another exact definition of
$\Lambda$ \cite{SVrev}, more in line with standard QCD conventions,
$\tilde\Lambda^3 =
{\mu^3 16\pi^2 \over 3c_2 g^2} \exp{-8\pi^2\over c_2 g^2},$
then the values of the gluino condensate in the Table~1 have to be adjusted
accordingly.}
\EQ{\Lambda^3 = \mu^3\frac1{g^2(\mu)} \exp\frac{2\pi i\tau(\mu)}{c_2}
\ .
\label{lexct}}
Here $\mu$ is the Pauli-Villars regulator mass and
$\tau$ is the usual complexified coupling incorporating both the gauge
coupling constant $g$ and the theta angle $\vartheta$:
\EQ{
\tau=\frac{\vartheta}{2\pi}+\frac{4\pi i}{g^2}\ .
\label{taudf}}

The paper is organized in the following way.
In \S\ref{sec:M2} we consider, in general terms, the effect of compactifying the
$\N=1$ gauge theory
on ${\mathbb R}^3 \times S^1$. \S\ref{sec:M22} discusses the various semi-classical
configurations that can contribute to the functional integral and
explains the relation between monopoles and instantons on the
cylinder. Regarding this point, our considerations in this paper
are purely field-theoretical; for an elegant D-brane discussion of
the $\SU(N)$ dynamics on ${\mathbb R}^3 \times S^1$ see
Refs.~\cite{DHKM,LY}. In \S\ref{sec:M23} we derive the form of the
superpotential in the low energy effective three-dimensional theory
generated by monopoles \eqref{sup}. It turns out that this potential
is precisely the affine Toda potential for a specific affine
algebra. For the simply-laced cases the affine algebra is the
untwisted affinization of the Lie algebra of the gauge group
while for the non-simply-laced cases it is
affine algebra whose Kac-Dynkin diagram is obtained from the untwisted affine
diagram with long roots changed to short roots, and {\it
vice-versa\/}. The affine algebras (in Kac's notation \cite{KAC}) are
listed in Table 2.
\begin{table}
\[
\begin{array}{ccc} \hline\hline
\Rsm\mbox{$\quad$ gauge group $\quad$} &\mbox{$\quad$ Lie algebra $\quad$}
&\mbox{$\quad$ affine Toda potential $\quad$} \\ \hline
\Rsm G (\text{Simply-laced}) & g & g^{(1)}\\
\Rsm \SO(2r+1) & b_r  & a^{(2)}_{2r-1}\\
\Rsm \USp(2r)  &  c_r &  d_{r+1}^{(2)}\\
\Rsm \G_2   &  g_2  &  d_4^{(3)}\\
\Rsm \F_4  &  f_4  &  e_6^{(2)}\\
\hline\hline
\end{array}
\]
\caption{\small The associated affine algebra.}
\end{table}
The Toda potential is in complete agreement with M(F)-theory considerations
\cite{Witten:1996bn,Katz:1997th,Vafa:1998vs}, although we will find
some additional pre-factors that feed into the calculation of the
gluino condensate in an essential way.
From the superpotential, we find that the number of
supersymmetric vacua is equal to dual Coxeter number of the gauge
group in complete agreement with the Witten index \cite{Windx}.
The values of the gluino condensate in each vacuum are then found and
the results are summarized in Table 1. For all classical groups
these are in agreement with the WCI results
of Refs.~\cite{NSVZtwo,Shifman:1988ia,Morozov,FP}.
In Appendix A we summarize our Lie algebra conventions and
Appendix B contains a brief discussion of the measure needed for integrating
over the collective coordinate space of fundamental monopoles.

\section{$\N=1$ gauge theory on the cylinder}\label{sec:M2}

In this section, we consider the effect of compactifying the pure $\N=1$ gauge theory
on a cylinder of radius $R$.
To this end, let us take $x_0$ to be periodic with period $2\pi R$.
We then impose
periodic boundary conditions on the gauge field and
gluino:\footnote{In our notation the four-dimensional indices run
over $m,n,\ldots=0,1,2,3$ while our three-dimensional indices run over
$\mu,\nu,\ldots=1,2,3$.}
\begin{equation}
v_m (x_\mu,x_0) = v_m (x_\mu,x_0+2\pi R) \ , \quad
\lambda (x_\mu,x_0) = \lambda (x_\mu,x_0+2\pi R) \ .
\label{ssbc}\end{equation}
Notice that the periodicity of the fermions preserves supersymmetry.

Smooth finite-action gauge fields on the cylinder
were classified in \cite{GPY}. In particular,
at finite radius instanton
configurations do not exhaust the set of semi-classical contributions.
The complete set of semi-classical configurations is
characterized by three pieces of data. Firstly, there is a topological
or instanton charge (or second Chern class) generalized from ${\mathbb R}^4$
to the cylinder:
\EQ{
k=\tfrac1{16\pi^2} \int_{{\mathbb R}^3\times S^1} d^4 x \ \tr\,v_{mn} \,
{}^*v^{mn}\ .
\label{tchcyl}
}
An important feature of the cylinder is that $k$ is not quantized in
integer units. However, when $k$ is an integer there are solutions
with action $S=8\pi^2k/g^2-ik\vartheta$ that, for scale size much
smaller than $R$, are identifiable as instantons of the
uncompactified theory. The second piece of data involves the Wilson loop of
the gauge field around the circle:
\EQ{
\oint_{S^1} dx_m\,v_m=\int_0^{2\pi R}dx_0\,v_0\equiv\varphi\ .
\label{ppr}
}
We will then define the VEV of $\varphi$ as the asymptotic value at spatial
infinity in ${\mathbb R}^3$:
\EQ{
\langle\varphi\rangle=\lim_{|x_\mu|\rightarrow\infty}
\boldsymbol\varphi\cdot\boldsymbol
H\ ,
\label{defavev}
}
where we have fixed a portion of the global gauge symmetry by
choosing the Wilson loop \eqref{defavev} to lie
within the Cartan subalgebra of the Lie algebra $g$
associated to the gauge group $G$.\footnote{Our Lie algebra conventions are
summarized in the Appendix A. We will denote $r={\rm rank}\,G$ vectors
in boldface.} This still leaves the freedom to
perform global gauge transformations from the Weyl group $W_g$ of $G$.

A non-zero value for $\langle\varphi\rangle$
acts as an adjoint-valued Higgs
field that generically
breaks the gauge group to its maximal abelian subalgebra $\U(1)^r$.
The classical moduli space ${\cal M}_{\rm cl}$, parameterized by the vector
$\langle\boldsymbol\varphi\rangle$, is the quotient
\EQ{
{\cal M}_{\rm cl}=\frac{{\mathbb R}^r}{2\pi\cdot\Lambda^*_W\rtimes W_g}\ ,
\label{modsp}
}
where $\Lambda_W^*$ is the co-weight lattice. the form of the quotient
is explained in the following way: we have already noted that fixing
$\langle\varphi\rangle$ to be in the Cartan subalgebra leaves the
freedom to perform global gauge transformations in the Weyl group. 
On top of this, theories with
$\langle\boldsymbol\varphi\rangle$ differing by $2\pi$ times any co-weight
vector are equivalent.
To see this last point, consider the following topologically non-trivial
gauge transformation
\begin{equation}
\sigma(x_0)=\exp \, \big({i x_0 \over R} \boldsymbol\omega^*\cdot\boldsymbol H\big) \ ,
\label{lgt}\end{equation}
for any co-weight $\boldsymbol\omega^*\in\Lambda^*_W$. This
transforms the component of the gauge field around the circle as
$v_0\to v_0+\boldsymbol\omega^*\cdot\boldsymbol H/R$, and hence $\langle\boldsymbol\varphi\rangle
\to\langle\boldsymbol\varphi\rangle+2\pi\boldsymbol\omega^*$.
The transformation \eqref{lgt} is periodic in the adjoint
representation of the gauge group\footnote{Because 
$\boldsymbol\alpha\cdot\boldsymbol\omega^*\in{\mathbb Z}$ for
{\it any\/} root $\boldsymbol\alpha$ and co-weight $\boldsymbol\omega^*$.}
and consequently in the pure gauge theory where all fields are
adjoint-valued $\langle\boldsymbol\varphi\rangle$ is identified with
$\langle\boldsymbol\varphi\rangle+2\pi\boldsymbol\omega^*$.

We will find it convenient to choose the VEV
$\langle\boldsymbol\varphi\rangle$ to lie in a ``fundamental cell''
\EQ{
0\leq\langle\boldsymbol\varphi\rangle\cdot\boldsymbol\alpha_i<2\pi\ ,\qquad i=1,\ldots
r,
\label{ordering}
}
where $\boldsymbol\alpha_i$ are the simple roots of
$g$.\footnote{Notice that this region is still an 
over-parameterization of the quotient \eqref{modsp}.}
The regions where
$\langle\boldsymbol\varphi\rangle\cdot\boldsymbol\alpha_i=0$, for some
set of $i$'s, correspond to submanifolds of ${\cal M}_{\rm cl}$
where a non-abelian subgroup of the gauge symmetry is restored.

The final piece of data arises from the fact that finite action
configurations can also carry three-dimensional magnetic charge. This
is an $r$-vector-valued quantity $\boldsymbol g$ in the charge space of the unbroken
$\U(1)^r$ abelian symmetry that can be defined via a surface
integral over the two-sphere at spatial infinity in ${\mathbb R}^3$ of
the magnetic field $B_\mu=\tfrac12\epsilon_{\mu\nu\rho}v_{\nu\rho}$:
\EQ{
-\tfrac1{2\pi}\int_{S^2}dS_\mu B_\mu
\equiv\boldsymbol g\cdot\boldsymbol H\ .
\label{magcharge}
}
The magnetic charge is subject to the usual generalization of the 
Dirac quantization rule \cite{Dirac,Weinberg:1982ev}
which requires 
that 
\EQ{
\boldsymbol g\in\Lambda_R^*\ ,
}
the co-root lattice of $g$.

Classically, the Wilson loop $\langle\boldsymbol\varphi\rangle$ is not
determined and so, as we have explained, there
is a moduli space ${\cal M}_{\rm cl}$ of inequivalent theories.
An important question is
whether this classical degeneracy persists in the quantum theory.
At this point, the behaviour depends crucially on whether one has
periodic or anti-periodic boundary conditions on the fermions. In the
latter---thermal---case, Ref.~\cite{GPY}
argued that non-trivial values of the asymptotic Wilson loop \eqref{defavev} are suppressed
in the infinite volume limit. Consequently, the classically flat directions
are lifted by thermal quantum corrections and the true vacuum of the theory is
$\langle\boldsymbol\varphi\rangle=0$. In this case, the configurations with magnetic
charges are not relevant, since they require non-vanishing VEV, 
and the semi-classical physics is described by instantons only.
Remarkably, for the theory on the cylinder, with periodic boundary
conditions on the fermions, the argument of \cite{GPY}
does not apply and, as we shall see in the following sections,
the opposite scenario ensues; namely:

(i) The semi-classical physics of the theory on the cylinder
is described by configurations of BPS
monopoles. There are $r+1$ types of ``fundamental'' monopole which
carry only four bosonic and two (adjoint)
fermionic zero modes. To those who are sufficiently initiated into
monopole calculus in gauge theories with arbitrary gauge group, this
will be a surprise: one would expected to have
only $r$ such monopoles (each with a magnetic charge equal to one of
the $r$ simple roots).
The additional monopole, needed to make up the
full complement of $r+1$ types,
is specific to the compactification on the cylinder
since, unlike the other, it is a non-trivial function of ``time''
$x_0$ \cite{LY,KL,LL,KvB}.
The magnetic charge of the new monopole is such that
when all $r+1$ types of monopoles are present with a specific
degeneracy, the magnetic charges
cancel and the resulting configuration carries only a unit of instanton
charge. Hence remarkably, instantons on the cylinder
can be understood as composite configurations
of monopoles \cite{Nahmtwo,Garland,LY,KL,LL,KvB}.

(ii) The classical moduli space of the gauge
theory on the cylinder \eqref{ppr}
is lifted in the quantum theory in a non-trivial way. The quantum
vacua correspond to a single point in ${\cal M}_{\rm cl}$ cell along with
an additional $c_2$-fold degeneracy, that has no counterpart in the
classical theory, and corresponds precisely to the expectations
based on a refined Witten index \cite{Windx} and the WCI counting
\cite{CD,Morozov}

\section{Semi-classical configurations}\label{sec:M22}

In the weak-coupling limit, the path integral is dominated by field
configurations which are of minimal action in each topological
sector. These configurations satisfy the four-dimensional self-dual, or anti-self-dual,
equations $v_{mn}=\pm^*v_{mn}$.
As we have explained there are two quantum numbers carried by semi-classical
configurations: the topological charge and the magnetic charge.

First of all, let us consider solutions which are independent of the
coordinate around the circle
$x_0$. These are simply BPS monopoles in the three-dimensional theory \cite{thm,polm,Bog,PS}
with the time direction taken to be along $x_0$.
Monopole solutions in a gauge theory with a simple gauge group $G$
can in turn be constructed out of the
$\SU(2)$ BPS monopole in the following way \cite{Weinberg:1982ev}.
The idea is to take a regular embedding $\SU(2)\subset G$,
associated to a positive root $\boldsymbol\alpha$ of
$G$:\footnote{Here,
$\boldsymbol\alpha^*=2\boldsymbol\alpha/\boldsymbol\alpha^2$ is the
co-root associated to $\boldsymbol\alpha$.}
\EQ{
\tau^1=\tfrac12(E_{\boldsymbol\alpha}+E_{-\boldsymbol\alpha})\
,\qquad\tau^2=\tfrac 1{2i}(E_{\boldsymbol\alpha}-E_{-\boldsymbol\alpha})
\ ,\qquad
\tau^3=\tfrac12\boldsymbol\alpha^*\cdot\boldsymbol H\ ,
\label{embedd}
}
which obey the $\SU(2)$ algebra
\EQ{
[\tau^a,\tau^b] = i \epsilon_{abc}\tau^c \ .
}
The monopole solution is then
\EQ{
v_0(x_\nu)=\Phi^c(v;x_\nu)\tau^c+\frac1{2\pi R}
\big(\langle\boldsymbol\varphi\rangle-
\tfrac12({\langle\boldsymbol\varphi\rangle\cdot\boldsymbol\alpha})\boldsymbol\alpha^*\big)\cdot
\boldsymbol H \ ,\qquad v_\mu(x_\nu)=v^c_\mu(v;x_\nu)\tau^c\ ,
\label{bpssunb}
}
where $\Phi^c(v;x_\nu)$ is the Higgs field and $v^c_\mu(v;x_\nu)$ are
the spatial components of the gauge field (in the
gauge $v_0=0$) of the $\SU(2)$ BPS monopole. The long distance
behaviour of this solution
\EQ{
\lim_{|x_\mu|\rightarrow\infty}\Phi^c(v;x_\nu)\tau^c=\frac v2
\boldsymbol\alpha^* \cdot \boldsymbol H
\ ,
\label{bhiggsvev}
}
where 
\EQ{
v=\frac{\boldsymbol\alpha \cdot \langle\boldsymbol\varphi\rangle}{2\pi R}\ .
\label{defv}
}
For this solution to be well defined, we must have $v>0$, which is automatic if
$\boldsymbol\alpha$ is a positive root and
$\langle\boldsymbol\varphi\rangle$ lies in the fundamental cell
\eqref{ordering}, in which case it
has magnetic charge, topological charge and action given by
\EQ{
\boldsymbol g=\boldsymbol\alpha^* \ ,
\qquad
k=
\boldsymbol\alpha^*\cdot{\langle{\boldsymbol\varphi\rangle} \over2\pi} \ ,
\qquad
S={4 \pi \over g^2}
\boldsymbol\alpha^*\cdot\langle{\boldsymbol\varphi\rangle}  \ .
\label{bmctoc}}
For completeness, we give the explicit solution for the $\SU(2)$ BPS monopole in ``hedgehog''
gauge
\AL{
v_\mu^c(v;x_\nu)& = \epsilon_{\mu \nu c} \frac{x_\nu}{\modx^2}
\left( 1 - \frac{v \modx}{\sinh v\modx} \right) \\
\Phi^c(v;x_\nu)& = \frac{x_c}{\modx^2} \left( v\modx \coth v\modx -1
\right) ,
}
The asymptotic value of the magnetic field of the hedgehog
solution, as $\modx \to \infty$, is
\EQ{
B_{\mu}^c  \to  - \frac{x_\mu x^c}{\modx^4} \ ,
%\qquad v_{\mu \nu}^c  \to  - \epsilon_{\mu \nu \rho} \ ,
\label{bhdbf}
}
while in unitary gauge
\EQ{
B_{\mu}^c  \to  - \frac{x_\mu}{\modx^3} \delta^{c3} \ , \qquad
B_{\mu} \equiv B_{\mu}^c \tau^c
\to - \frac{x_\mu}{2\modx^3} \boldsymbol\alpha^* \cdot H \ .
\label{bugbf}}

However, these $x_0$-independent solutions do not exhaust the set of
solutions with a given magnetic charge
$\boldsymbol\alpha^*$ \cite{LY}. A whole tower of other
solutions which {\it are\/} $x_0$ dependent can be generated in the following way. 
First of all, we start with the
solution \eqref{bpssunb} with $\langle\boldsymbol\varphi\rangle$ lying
in the fundamental cell \eqref{ordering}. We then write down the same
solution with a shifted VEV
$\langle\boldsymbol\varphi'\rangle=\langle\boldsymbol\varphi\rangle+
\pi n\boldsymbol\alpha^*$, where $n\in{\mathbb Z}$. For this solution
to be well defined we must have 
\EQ{
v'=\frac{\boldsymbol\alpha\cdot\langle\boldsymbol\varphi'\rangle}{2\pi R}
=\frac{\boldsymbol\alpha\cdot\langle\boldsymbol\varphi\rangle}{2\pi R}+\frac
nR>0\ ,
\label{iioo}
}
For $\boldsymbol\alpha=\boldsymbol\alpha_i$, a simple root, \eqref{ordering} implies
that $n\geq0$. Acting on the
solution with the (non-periodic) gauge transformation
\EQ{
V_n(x_0)=\exp\big(\frac{inx_0}{2R}\boldsymbol\alpha^*\cdot\boldsymbol H\big)\ ,
\label{largeg}
}
has the effect of restoring the VEV $\langle\boldsymbol\varphi\rangle$ to
its original value. The new solution is then given by
\SP{
v_0(x_\nu)&=\Phi^c(v+n/R;x_\nu)
\tilde\tau^c+\frac1{2\pi R}
\big(\langle\boldsymbol\varphi\rangle-
\tfrac12({\langle\boldsymbol\varphi\rangle\cdot\boldsymbol\alpha}+2\pi n)\boldsymbol\alpha^*\big)\cdot
\boldsymbol H \ ,\\
v_\mu(x_\nu)&=v^c_\mu(v+n/R;x_\nu)\tilde\tau^c\ ,
\label{newbpssunb}
}
where $v$ is given as in \eqref{defv}
and the $\SU(2)$ generators are conjugated with $V_n(x_0)$:
\EQ{
\tilde\tau^c=V_n(x_0)\tau^cV_n(x_0)^{-1}\ .
}
Notice although $V_n(x_0)$ is not a periodic gauge transformation
the generators $\tilde\tau^c$ {\it are\/} periodic
functions of $x_0$.
The solution \eqref{newbpssunb} has the same magnetic charge as
\eqref{bpssunb}, but the topological charge is
$k=\boldsymbol\alpha^*\cdot\langle\boldsymbol\varphi\rangle/2\pi+n$.
This solution can be interpreted as a composite configuration of the
original monopole plus an instanton of charge $n$.

However, there are also towers of solutions of the self-dual equations that
have a magnetic charge equal to some {\it negative\/} root \cite{LY}. We
should emphasize that these solutions are {\it not\/} anti-monopoles which
would satisfy the {\it anti\/}-self-dual equations. To construct these
solutions we can start with our solution \eqref{bpssunb} with
$\langle\boldsymbol\varphi\rangle$ lying in the fundamental cell.
We now define a new solution with a VEV
$\langle\boldsymbol\varphi'\rangle=
\sigma_{\boldsymbol\alpha}(\langle\boldsymbol\varphi\rangle)
+\pi n\boldsymbol\alpha^* $, 
where $\sigma_{\boldsymbol\alpha}$ is the Weyl reflection in
$\boldsymbol\alpha$. For the solution to be well defined
we must have 
\EQ{
v'=\frac{\boldsymbol\alpha\cdot\langle\boldsymbol\varphi'\rangle}{2\pi R}
=-\frac{\boldsymbol\alpha\cdot\langle\boldsymbol\varphi\rangle}{2\pi R}+
\frac nR>0\ .
\label{iiooo}
}
For $\boldsymbol\alpha=\boldsymbol\alpha_i$,  a simple root, this
means $n>0$. To re-install
the original VEV, we then perform a Weyl reflection in $\boldsymbol\alpha$ and
the gauge transformation \eqref{largeg}. The resulting
solution is
\SP{
v_0(x_\nu)&=\Phi^c(n/R-v;x_\nu)\tilde\tau^c
+\frac1{2\pi R}
\big(\langle\boldsymbol\varphi\rangle-
\tfrac12(-{\langle\boldsymbol\varphi\rangle\cdot\boldsymbol\alpha}
+2\pi n)\boldsymbol\alpha^*\big)\cdot
\boldsymbol H \ ,\\
v_\mu(x_\nu)&=v^c_\mu(n/R-v;x_\nu)\tilde\tau^c\ ,
\label{newbpssunc}
}
where $v$ is given in \eqref{defv} and
the $\SU(2)$ generators are now conjugated with
$V_n(x_0)\sigma_{\boldsymbol\alpha}$:
\EQ{
\tilde\tau^c=V_n(x_0)\sigma_{\boldsymbol\alpha}\tau^c\sigma_{\boldsymbol\alpha}V_n(x_0)^{-1}
\ .
}
It can be easily verified that this solution is again periodic in $x_0$.
The resulting solution has magnetic charge $-\boldsymbol\alpha^*$ and
topological
charge $k=-\boldsymbol\alpha^*\cdot\langle\boldsymbol\varphi\rangle/2\pi+n$.

It will be important for later to determine the number of adjoint
fermion, or gluino, zero modes of these monopole solutions.
Each classical solution has at
least two adjoint fermion zero modes protected by
supersymmetry.
These modes can be generated from the purely bosonic
solution by acting with the generators of supersymmetry that do not
leave the configuration invariant.
This gives the universal expression
for these supersymmetric modes
\EQ{
\lambda_\alpha=\sigma^{mn\beta}_\alpha\xi_\beta v_{mn} \ ,
\label{lss}
}
where $v_{mn}$ is the field strength.
For future reference we give the long-distance behaviour of the supersymmetric
fermion zero modes \eqref{lss} of our fundamental monopole solutions
\eqref{bpssunb} and \eqref{newbpssunc} with $n=1$:
\EQ{\lambda_\alpha = \sigma^{mn\phantom{\alpha}\beta}_{\phantom{mn}\alpha}
\xi_\beta v_{mn} = -2 (\sigma^{\nu} \xi)_\alpha B_\nu  \to  4 \pi (
{\cal S}_{\rm F} \xi)_\alpha \boldsymbol\alpha^* \cdot \boldsymbol H \ ,
\label{bssfld}}
where ${\cal S}_{\rm F}(x)=\sigma_\mu x_\mu/(4\pi|x_\mu|^3)$
is the massless fermion propagator in
three dimensions.

Solutions
with only the supersymmetric zero modes have four associated bosonic
zero modes which correspond to moving the centre-of-mass of the
monopole in ${\mathbb R}^3$ as well as performing global gauge
rotations by
$\exp( \tfrac i2\Omega\ \boldsymbol\alpha^*\cdot\boldsymbol H)$. Hence these
solutions are special in that they are elementary or ``fundamental'':
the other solutions have
additional moduli that correspond to pulling the configuration apart
into their fundamental constituents.

As might have been expected there are $r$ solutions of the form \eqref{bpssunb}
where $\boldsymbol\alpha$ is a simple root $\boldsymbol\alpha_i$ lying
at the bottom of the more general tower of solutions
\eqref{newbpssunb}. 
This gives us $i=1,\ldots,r$ fundamental monopole solutions
with two adjoint-valued fermion zero modes, magnetic
charge $\boldsymbol\alpha^*_i$, and the topological charge
$k=\boldsymbol\alpha^*_i\cdot\langle\boldsymbol\varphi\rangle/2\pi$. Solutions
higher in the tower, with $n>0$, have $2(1+nc_2)$
fermion zero modes
\cite{LY,KL}, as we expect for a configuration of a fundamental
monopole and $n$ instantons.
In addition to these $r$ fundamental monopoles,
there is one other solution that
is fundamental \cite{LY,KL}. This is solution
which has a negative magnetic charge equal to the lowest root
$\boldsymbol\alpha_0^*\equiv\boldsymbol\alpha_0$
(although the solution, as we explained above is
{\it not\/} an anti-monopole) lying in the second tower
\eqref{newbpssunc} with $n=1$ and hence with topological charge
$k=-\boldsymbol\alpha_0^*\cdot\langle\boldsymbol\varphi\rangle/2\pi+1$.

Since $\sum_{i=0}^r k_i^*
\boldsymbol\alpha_i^*=0$,\footnote{Here $k_i^*$ are the dual ``Kac labels'', or co-marks,
defined in Appendix A.} the
quantum numbers of the solutions suggest that a pure instanton
solution, carrying zero magnetic charge and unit topological charge, is
a composite configuration with $k_i^*$ fundamental $\boldsymbol\alpha_i$
monopoles, for each $i=0,\ldots,r$. 
This turns out to be the case \cite{LY,KL} and the resulting configuration has
exactly $2c_2$ ($4c_2$) exact fermionic (bosonic)
zero modes as expected for a singly-charged
instanton with gauge group $G$.

\section{Monopole contributions to the superpotential}\label{sec:M23}

In this section, we will explain how the fundamental monopoles described in the
last section lift the classical degeneracy of the theory
parametrized by the asymptotic value of the Wilson loop
$\langle\boldsymbol\varphi\rangle$ \eqref{defavev}.
The idea is to consider the low energy three-dimensional effective
theory corresponding to the massless abelian components of the fields formed by
integrating out all the massive fields.

For this analysis to hold we must first
assume there is no root $\boldsymbol\alpha$ such that
$\langle\boldsymbol\varphi\rangle\cdot\boldsymbol\alpha=0$, so that the unbroken
gauge group is maximally abelian $\U(1)^r$. We will also assume that the Wilson line VEV
$\langle\boldsymbol\varphi\rangle$ lies in the fundamental region \eqref{ordering}.
After that we can integrate out
(1) all non-abelian fields on ${\mathbb R}^3 \times S^1$, and (2)
all the massive
Kaluza-Klein modes on $S^1$, {\it i.e.\/}~the modes with non-zero Matsubara frequency
$\omega_m=m/R$, to flow to the abelian theory on ${\mathbb R}^3$.
We emphasize that the periodicity in
$\langle\boldsymbol\varphi\rangle\sim\langle\boldsymbol\varphi\rangle+2\pi\boldsymbol\omega^*$,
$\boldsymbol\omega^*\in\Lambda_W^*$,
is a property of the full microscopic theory but not of the low-energy
theory on ${\mathbb R}^3$.
Indeed, the large gauge transformation \eqref{lgt} is $x_0$-dependent
and has the effect of mixing up the massless and massive Kaluza-Klein modes.

The fields of the low energy theory
consist of the Wilson loop $\boldsymbol\varphi$, {\it i.e.\/}~the 
component $v_0$ of the gauge field averaged over the circle,
along with $r$ massless photons corresponding to the
components of $v_\mu$ in the Cartan subalgebra of the gauge
group. Along with these bosonic fields there are superpartners corresponding
to the abelian components of the gluino.

It turns out to be convenient to use the fact that massless abelian
gauge fields in three dimensions can be eliminated in favour of scalar
fields by a duality transformation. To construct the classical
effective action, we start with the
action of the pure gauge theory in four
dimensions and dimensionally reduce to three dimensions
keeping only the abelian components of the fields. From \eqref{ppr},
the component $v_0$ of the four-dimensional gauge field
is replaced by $\boldsymbol\varphi\cdot\boldsymbol H/(2\pi R)$ and the resulting
three-dimensional effective action is\footnote{It is
useful to notice that in our normalization $\tr(\boldsymbol a\cdot
\boldsymbol H \ 
\boldsymbol b\cdot\boldsymbol H)=\boldsymbol a\cdot\boldsymbol b$.}
\EQ{
S_{\text{cl}}=\frac{2\pi R}{g^2}\int d^3x\,
\Big\{\frac1{4\pi^2R^2}(\partial_\mu
\boldsymbol\varphi)^2-\tfrac12
(\boldsymbol v_{\mu\nu})^2+2i\boldsymbol{\bar\lambda}\cdot\bar\sigma_\mu
{\cal D}_\mu\boldsymbol\lambda\Big\}-\frac{i\vartheta}{8\pi^2}\int d^3x\,
\epsilon_{\mu\nu\rho}\partial_\mu\boldsymbol\varphi\cdot\boldsymbol v_{\nu\rho}\ .
\label{dimract}
}
In order to construct the dual description of the three-dimensional
gauge field one adds a new term to the action involving a
field $\boldsymbol\sigma$ which serves
as a Lagrange multiplier for the Bianchi identity constraint:
\EQ{
S_{\text{cst}}=-\tfrac i{4\pi}\int
d^3x\,\epsilon_{\mu\nu\rho}\partial_\mu
\boldsymbol\sigma\cdot\boldsymbol v_{\nu\rho}=-\tfrac i{2\pi}\int_{S^2}dx_\mu
\,\boldsymbol\sigma\cdot\boldsymbol B_\mu\ .
\label{surface}
}
The abelian field strength $\boldsymbol v_{\mu\nu}$ can now be integrated-out
of the path integral as a Gaussian field to obtain the
classical effective action, whose bosonic part is
\EQ{
S_{\text{cl}}^{\rm bos}=
\frac1{2\pi R}\int d^3x\,
\Big\{\frac1{g^2}(\partial_\mu
\boldsymbol\varphi)^2+\frac{g^2}{16\pi^2}\big(
\partial_\mu\boldsymbol\sigma+\frac{\vartheta}{2\pi}\partial_\mu\boldsymbol\varphi\big)^2\Big\}\ .
}
This can be written compactly in terms of the single complex field
\EQ{
\boldsymbol z=i(\tau\boldsymbol\varphi+\boldsymbol\sigma) .
\label{defcz}
}
as
\EQ{
S_{\text{cl}}^{\rm bos}
=\frac{1}{8\pi^2 R}\int d^3x\, \frac1{{\rm Im}\tau}\partial^\mu
\boldsymbol z^+\cdot\partial_\mu\boldsymbol z\ .
\label{clbos}
}
We have eliminated the $r$ massless
photons in favour of an $r$-vector scalar field
$\boldsymbol\sigma$. Notice that since the
magnetic charge $\boldsymbol g$ is quantized in the co-root lattice
it follows from \eqref{surface} that $\boldsymbol\sigma$ is 
physically equivalent to $\boldsymbol\sigma+2\pi\boldsymbol\omega$ for any weight
$\boldsymbol\omega\in\Lambda_W$. Once again, we also have the freedom to
perform Weyl reflections and so $\boldsymbol\sigma$ is valued in the quotient
\EQ{
\frac{{\mathbb R}^r}{2\pi\cdot\Lambda_W\rtimes W_g}\ ,
\label{nmodsp}
}
to compare with $\boldsymbol\varphi$ which is valued in the slightly
different quotient
\eqref{modsp}. Obviously these spaces are the same for the
simply-laced groups.

The fact that both (real) scalar fields $\boldsymbol\varphi$ and
$\boldsymbol\sigma$ can be amalgamated into a single complex field $\boldsymbol z$ is
no coincidence. Since the original four-dimensional theory was ${\cal N}=1$ supersymmetric, the
effective theory written in terms of the bosonic fields
$\boldsymbol\varphi$ and $\boldsymbol\sigma$, along with the abelian components of the
gluino $\boldsymbol\lambda_\alpha$, must form a representation of
four-dimensional ${\cal
N}=1$ supersymmetry\footnote{corresponding to $\N=2$ in three
dimensions.} which must be a chiral superfield since
we have taken the dual of all the vector fields.
In particular the bosonic fields must be expressible in terms of a
single complex field as we have found in \eqref{defcz}.

The abelian gluino fields simply complete \eqref{clbos} to the
supersymmetric invariant expression written in terms of the
dimensional reduction of a four-dimensional ${\cal
N}=1$ chiral superfield $\boldsymbol X$ with scalar component $\boldsymbol z$ and fermionic
component $\boldsymbol\lambda_\alpha$, the abelian component of the gluino.
The supersymmetric version of \eqref{clbos} written in superspace is
then 
\EQ{
S_{\text{cl}}=\frac1{8\pi^2 R}\int d^3x\, \frac1{{\rm Im}\tau}
\boldsymbol X^+\cdot\boldsymbol
X\Big|_{\theta\theta\bar\theta\bar\theta}\ .
\label{sscleff}
}

Quantum effects can modify the classical expression
\eqref{sscleff}. However, modifications must be consistent with ${\cal
N}=1$ supersymmetry. As long as we are at a generic point in the
classical moduli space, we expect to be able to integrate out all the
massive fields to be left with an effective theory in terms of the
superfield $\boldsymbol X$. The most general possible low energy effective
action, {\it i.e.\/}~involving at most two derivatives or four fermions, is
\EQ{
S_{\text{eff}}=\int d^3x\,\Big\{
{\cal K}(\boldsymbol X,\boldsymbol
X^+)\Big|_{\theta\theta\bar\theta\bar\theta}
+{\cal W}(\boldsymbol X)\Big|_{\theta\theta}
+\overline{\cal W}(\boldsymbol X^+)\Big|_{\bar\theta\bar\theta}\Big\}\ ,
\label{ssqeff}
}
which involves an arbitrary $D$-term ${\cal K}(\boldsymbol X,\boldsymbol
X^+)$ as well as a superpotential ${\cal W}(\boldsymbol X)$. It is the
superpotential that is responsible for lifting the classical degeneracy
and which we must determine.

In the classical theory \eqref{sscleff} the superpotential vanishes
identically. Quantum corrections will modify the theory in a
complicated way depending on the couplings. However, the
superpotential, by the standard arguments \cite{Seiberg93,
Intriligator:1996au,SWthree},
must be holomorphic
in the fields $\boldsymbol X$ and the complexified coupling $\tau$. In
particular, up to the overall factor,\footnote{This appears when the fields
are not canonically normalized. In our case scalar fields arise from the
Wilson line and the dual photon, which are dimensionless. This leads to
an overall factor of $R/g^2$ in Eq.~\eqref{sup} below.}
the superpotential
can only depend on $R$ through the running of $\tau$
via the dimensionless quantity $R|\Lambda|$, where
$\Lambda$ is the usual Pauli-Villars scale of strong coupling effects in
the pure gauge theory in ${\mathbb R}^4$. 
We intend to compute the superpotential at weak-coupling, for which
$R\ll|\Lambda|^{-1}$ and the VEV of the effective Higgs field \eqref{defv}
is large and a semi-classical analysis
should be reliable. 
In this regime the superpotential will receive contributions from
the minimal action configurations in each topological sector which
have exactly two gluino zero modes;
in other words from the $r+1$ fundamental monopoles
described in the last section. As usual holomorphy then forbids 
any perturbative corrections to the semi-classical contributions
and, as a consequence, fixes the $R$ dependence, a fact that ultimately will
allow us to take $R$ to be large.

In the presence of the dual photon field $\boldsymbol\sigma$, the
action of the fundamental monopole associated to the root
$\boldsymbol\alpha_j$, $j=0,\ldots,r$, is given in
terms of the VEV of the scalar field $z$ by
\EQ{
S_j=-2\pi i\tau\delta_{j0}
-i\tau\boldsymbol\alpha_j^*\cdot\langle\boldsymbol\varphi\rangle
-i\boldsymbol\alpha_j^*\cdot\langle\boldsymbol\sigma\rangle\equiv
-2\pi i\tau\delta_{j0}-\boldsymbol\alpha_j^*\cdot\langle\boldsymbol z\rangle\ .
\label{monact}
}
Here $\tau$ is the complexified coupling \eqref{taudf}.

We determine the form of the superpotential by
calculating the monopole contribution to the large distance behaviour
of the correlator of two components of the massless gluino field
\EQ{
\VEV{\boldsymbol\lambda_\alpha(x) \otimes
\boldsymbol\lambda_\beta(0)}\ .
}
In the background of the $\boldsymbol\alpha_j$ monopole, only the component
$\boldsymbol\lambda_\alpha\propto\boldsymbol\alpha_j$ is non-trivial; in fact
from \eqref{bssfld} one finds the long-distance behaviour to be
\EQ{
\boldsymbol\lambda^{\sst LD}_\alpha(x) \ = \
4\pi\boldsymbol\alpha_j^*{\cal S}_{\rm
F}(x-a)_\alpha^{\ \gamma}\xi_\gamma\ ,
}
where ${\cal S}_{\rm F}(x)=\sigma_\mu x_\mu/(4\pi|x_\mu|^3)$
is the massless fermion propagator in
three dimensions, $a_\mu$ is the position of the monopole in ${\mathbb
R}^3$ and $\xi_\alpha$ are the Grassmann collective
coordinates corresponding to the two supersymmetric zero modes.

In order to evaluate the contribution to the superpotential from the
monopole, we need the measure for
integrating over the moduli space of the monopole 
derived in the Appendix B.
A fundamental monopole has a moduli space that is
parametrized by $a_\mu$, the
position in ${\mathbb R}^3$ and by the $\U(1)$ phase angle
$0\leq\Omega\leq2\pi$. Along with this, there are two Grassmann collective
coordinates $\xi_\alpha$, corresponding to the two supersymmetric
zero modes. From Eq.~\eqref{bmeas} the measure is
\EQ{
\int d\mu_{\rm mon}^{(j)}=\frac{2}{\boldsymbol\alpha_j^2}\frac{\mu^3R}{g^2}e^{-S_j}
\int d^3 a\,d\Omega\,d^2\xi\ .
\label{msst}
}
Performing the integrals over the phase angle and the Grassmann
collective coordinates, we find that
\EQ{
\VEV{\boldsymbol\lambda_\alpha(x) \otimes
\boldsymbol\lambda_\beta(0)}
=\frac{2^6\pi^3\mu^3R}{g^2\boldsymbol\alpha_j^2}
\boldsymbol\alpha_j^*\otimes\boldsymbol\alpha_j^*
e^{2\pi i\tau\delta_{j0}+\boldsymbol\alpha_j^*\cdot\langle\boldsymbol
z\rangle}
\int d^3a\, {\cal S}_{\rm F}(x-a)_{\alpha}^{\  \gamma}
{\cal S}_{\rm F}(a)_{\beta \gamma}\ ,
\label{fermc}
}
Amputating this correlator we find the associated
vertex in the effective action:
\EQ{
\Big(\frac{2\pi R}{g^2}\Big)^2 \
\frac{2^5\pi^3\mu^3R}{g^2
\boldsymbol\alpha_j^2} \ e^{2\pi i\tau\delta_{j0}
+\boldsymbol\alpha_j^*\cdot\langle\boldsymbol
z\rangle} \ (\boldsymbol\alpha_j^*\cdot\boldsymbol\lambda)^2\ .
\label{vertex}
}
In the above, the numerical factor in the bracket reflects our
normalization for the kinetic term of $\boldsymbol\alpha\cdot\boldsymbol\lambda$
which follows from \eqref{dimract}.
The vertex \eqref{vertex} is generated by a term in the effective potential of the
form\footnote{In order to get the correct numerical factor, notice
that the fermionic component of $\boldsymbol z$ and the gluino $\boldsymbol\lambda$
are related via $\boldsymbol\psi=2^{5/2}\pi^2 g^{-2}R\,\boldsymbol\lambda$.
This follows from the fact \cite{DKMTV}
that the superpartner of $\boldsymbol v_0$ is
$(\boldsymbol\lambda+\boldsymbol{\bar\lambda})/\sqrt{2}$.
}
\EQ{
\frac{4\pi\mu^3 R}{g^2\boldsymbol\alpha_j^2}e^{2\pi i\tau\delta_{j0}+
\boldsymbol\alpha_j^*\cdot\boldsymbol X}\ .
}
Hence, summing over the the effects of all $r+1$ fundamental monopoles we
deduce that the monopole-generated superpotential of the theory is
\EQ{
{\cal W}_{\rm mono}
(\boldsymbol X)=\frac{2\pi\mu^3 R}{g^2}\Big(\sum_{j=1}^{r}\frac2{\boldsymbol\alpha_j^2}
e^{\boldsymbol\alpha^*_{j}\cdot\boldsymbol X}+\frac2{\boldsymbol\alpha_0^2}
e^{2\pi i\tau+\boldsymbol\alpha_0^*\cdot\boldsymbol X}\Big)\ .
\label{sup}
}
This is an affine Toda potential for an associated affine
algebra. Notice that to give the usual expression for a Toda
potential one can remove the pre-factors $2/\boldsymbol\alpha_j^2$ by
a shift in the field:
\EQ{
\boldsymbol X\to\boldsymbol X+\sum_{j=1}^r\ln(\boldsymbol\alpha_j^2/2)\boldsymbol\omega_j+
\frac{\boldsymbol\rho}{c_2}\big(2\pi i\tau-\sum_{j=0}^r\ln(\boldsymbol\alpha_j^2/2)\big)\ ,
}
where $\boldsymbol\rho=\sum_{j=1}^r\boldsymbol\omega_j$ is the Weyl vector.
For the simply-laced groups, the associated affine algebra is the
untwisted affinization of the original Lie algebra, $g^{(1)}$ in Kac's
notation \cite{KAC}. While for the 
non-simply-laced groups the corresponding affine algebra is twisted in
the way described in the Table~2. In these cases, the Kac-Dynkin diagram
of the affine algebra is obtained from the Kac-Dynkin diagram of the
untwisted affinization $g^{(1)}$ by changing long roots into short
roots, and {\it vice-versa\/}. In Kac's notation \cite{KAC} this 
leads to the twisted affinization of a different algebra.
The same superpotential has been deduced from entirely different considerations
involving M theory compactified on certain 8 dimensional manifolds
\cite{Witten:1996bn,Katz:1997th,Vafa:1998vs,
Gomez:1997mq,Gomez:1997}, although the $2/\boldsymbol\alpha_j^2$ pre-factors, that
we shall find crucial in order to get results for the gluino
condensate that agree with other calculations, are absent.
It is also interesting that the integrable systems related to
the Toda
potentials that we have found above are precisely those that appear
in the ``Seiberg-Witten theory'' of the $\N=2$ gauge
theory with the same gauge group in four dimensions
\cite{MW,TJH}. Naturally this is no accident since the $\N=1$ theory
can be obtained from the $\N=2$ theory by soft breaking mass terms.

Importantly, although we have calculated the superpotential in the
limit $R\ll|\Lambda^{-1}|$, at weak coupling, there can be no additional
dependence on $R$ and the result can be
continued to any $R$, and in particular to the decompactification
limit \cite{SWthree,Aharony:1997bx}. 

One may wonder how the
superpotential relates to that calculated in \cite{AHW}
for the three-dimensional ${\cal N}=2$ supersymmetric gauge
theory. The way that this superpotential arises from the $R\to0$
limit of our superpotential is explained in a slightly different
context in \cite{SWthree}. The point
is that to take the three-dimensional limit, one should take it in
such a way that the three-dimensional gauge coupling, which is
classically given by $g_3^2=g^2/(2\pi R)$, is fixed. In other words,
as $R\to0$, we
should simultaneously take the limit $g\to0$ in the superpotential
\eqref{sup}. In this limit, the additional term corresponding to the
affine root is removed to give
\EQ{
{\cal W}_{\text{3-d}}=\frac{\mu^3}{g_3^2}\sum_{j=1}^{r}\frac2{\boldsymbol\alpha_j^2}
e^{\boldsymbol\alpha_j^*\cdot\boldsymbol X}\ .
\label{tdsp}
}
In other words, in this limit, the affine Toda potential becomes the Toda potential
for a non-affine algebra. This is what one expects because the
affine term in the superpotential on the cylinder is generated by the
additional monopole solution that only exists on the cylinder and not
in ${\mathbb R}^3$. The genuinely three-dimensional
superpotential \eqref{tdsp} is the generalization of that of
\cite{AHW} from $\SU(2)$ to arbitrary gauge group. Just as in the $\SU(2)$ case, it does
not have a stationary point and therefore the theory does not have a vacuum
state.

\section{Vacuum structure and the gluino condensate}\label{sec:M24}

The superpotential \eqref{sup}
gives rise to a number of supersymmetric vacua
which satisfy
\EQ{
\frac2{\boldsymbol\alpha_j^2}e^{\boldsymbol\alpha_j^*\cdot\boldsymbol X}=
\frac2{\boldsymbol\alpha_0^2} k_j^*e^{2\pi
i\tau}e^{\boldsymbol\alpha_0^*\cdot\boldsymbol X}\ ,
}
for $j=1,\ldots,r$. Writing $\boldsymbol X=\sum_{j=1}^r a_j
\boldsymbol\omega_j$ we have
\EQ{
e^{a_j}=\frac{k_j^*\boldsymbol\alpha_j^2e^{2\pi i\tau}}{2\kappa}\ ,
}
where $\kappa=e^{\sum_{j=1}^r a_j k_j^*}$ is determined
self-consistently as the solution of the equation
\EQ{
\kappa^{c_2}=e^{2\pi i(c_2-1)\tau}\prod_{j=0}^r\Big(\frac{k_j^*\boldsymbol\alpha_j^2}2\Big)^{k_j^*}\ .
\label{lsol}
}
There are consequently $c_2$ supersymmetric ground states given by the
$c_2$ roots of \eqref{lsol} which are related by $\boldsymbol X\to\boldsymbol X+2\pi
i\boldsymbol\rho/c_2$.\footnote{The vector
$\boldsymbol\rho=\sum_{j=1}^r\boldsymbol\omega_j$ is the Weyl vector
and recall that $\boldsymbol X$ is identified with $\boldsymbol
X+2\pi i\boldsymbol\rho$ as a consequence of the fact that $\boldsymbol\sigma$ is
identified with $\boldsymbol\sigma+2\pi\boldsymbol\rho$, 
since $\boldsymbol\rho\in\Lambda_W$.} These vacua
correspond to a fixed  value of $\boldsymbol\varphi$:
\EQ{
\boldsymbol\varphi=\Big(\frac{2\pi}{c_2}+\frac{g^2}{4\pi}\ln|\kappa|
\Big)\boldsymbol\rho-\frac{g^2}{4\pi}
\sum_{j=1}^r\ln(k_j^*\boldsymbol\alpha_j^2/2)\boldsymbol\omega_j\ .
}
and $c_2$ values of $\boldsymbol\sigma$ given by
\EQ{
\boldsymbol\sigma=-\frac{\vartheta}{2\pi}\boldsymbol\varphi+\frac{\vartheta+2\pi u}
{c_2}\boldsymbol\rho\ ,
}
where $u=1,2,\ldots,c_2$. Notice that as expected the $c_2$ vacua are
related by $\vartheta\to\vartheta+2\pi$. 

The value of the superpotential in one of the vacua is
\SP{
\langle{\cal W}_{\rm mono}\rangle&=\frac{2\pi\mu^3R}{g^2}\cdot\frac{e^{2\pi
i\tau}c_2}\kappa
=\frac{2\pi\mu^3R}{g^2}\cdot\frac{c_2e^{2\pi i\tau/c_2+2\pi
iu/c_2}}{\prod_{j=0}^r(k_j^*\boldsymbol\alpha_j^2/2)^{k_j^*/c_2}} \\
&=2\pi R\Lambda^3\cdot\frac{c_2 e^{2\pi
iu/c_2}}{\prod_{j=0}^r(k_j^*\boldsymbol\alpha_j^2/2)^{k_j^*/c_2}}\ ,
}
where in the final expression
we have eliminated the Pauli-Villars mass scale $\mu$ in favour
of the Lambda parameter using the exact
relation \eqref{lexct}.
The value of the gluino condensate in each vacuum can the extracted by
using the
general relation
\EQ{
\VEV{\frac{\tr\,\lambda^2}{16\pi^2}}=b_0^{-1}\Lambda\frac{\partial}
{\partial\Lambda}\vev{{1\over 2\pi R}{\cal W}_{\rm mono}
}\ ,
\label{gcads}
}
adapted to the three-dimensional
superpotential. The first coefficient of the
beta-function is $b_0=3c_2$ giving
\EQ{
\VEV{\frac{\tr\lambda^2}{16\pi^2}}=
\frac{\Lambda^{3}e^{2\pi iu/c_2}}{\prod_{j=0}^r(k_j^*\boldsymbol\alpha_j^2/2)^{k_j^*/c_2}}\ .
\label{gcond}
}

The gluino condensate can also be evaluated directly without having to
rely on the identity \eqref{gcads}. The idea is to consider the
fundamental monopole contributions to the one-point function
$\vev{\frac{\tr\,\lambda^2}{16\pi^2}}$ in a given vacuum, say the
$u^{\rm th}$. The contribution of the
$\boldsymbol\alpha_j$ monopole to the condensate in this vacuum is
\EQ{
\VEV{\frac{\tr\,\lambda^2}{16\pi^2}}_{j\text{-mono}}
=\int  d\mu_{\rm
mon}^{(j)}\,\frac{\tr\,\lambda^2}{16\pi^2}\Big|_{j\text{-mono}}\
.
\label{umonc}
}
To evaluate \eqref{umonc}, we can use the normalization of the adjoint
fermion zero modes from Ref.~\cite{DKMTV}
\EQ{
\int d^3a\,
d^2\xi\frac{\tr\,\lambda^2}{16\pi^2}\Big|_{j\text{-mono}}
=\frac{g^2{\rm Re}\,S_j}{8\boldsymbol\alpha_j^2\pi^3 R}\ .
}
Computing the remaining integral over the phase angle gives
\EQ{
\VEV{\frac{\tr\,\lambda^2}{16\pi^2}}_{j\text{-mono}}
=\frac{\mu^3{\rm Re}\,S_j}{4\pi^2\boldsymbol\alpha_j^2}e^{-S_j}\ .
}
In the supersymmetric vacua
\EQ{
S_j=-2\pi i\tau-\ln\Big(\frac{k_j^*\boldsymbol\alpha_j^2}{2\kappa}\Big)
}
and so inserting the value for $\kappa$ in \eqref{lsol} we have
\SP{
\VEV{\frac{\tr\,\lambda^2}{16\pi^2}}_{j\text{-mono}}
=\frac{k_j^*\Lambda^{3}e^{2\pi iu/c_2}}{c_2}
\cdot\frac1{\prod_{j=0}^r(\boldsymbol\alpha_j^2k_j^*/2)^{k_j^*/c_2}}\ .
}
Summing over the contributions from the $r+1$
fundamental monopoles gives \eqref{gcond}.

We conclude the section with the observation that
in the supersymmetric vacua the $r+1$ fundamental
monopoles have equal topological charge 
\EQ{
\boldsymbol\alpha_j\cdot\frac{\langle\boldsymbol\varphi\rangle}{2\pi}=
1-\boldsymbol\alpha_0\cdot\frac{\langle\boldsymbol\varphi\rangle}{2\pi}=\frac
1{c_2}\ ,
}
(for $j=1,\ldots,r$) independent of $j$. In addition, as we have
discussed in \S\ref{sec:M22} the configuration
which becomes the singly-charged instanton in the uncompactified
theory is obtained by considering a multi-monopole solution which
consists of $k_j^*$ of 
the $j^{\rm th}$ fundamental monopole. In this very precise sense they
realize the old dream of thinking of the instanton in terms of a set of
constituents, or instanton quarks
\cite{Belavin:1979fb,Fateev:1979dc,Berg:1979uq,Callan:1978gz,Osborn}.
It was anticipated that the instanton quarks
would cause, or at least
play a major r\^ole in, confinement. In the theory on the cylinder
this old idea again receives confirmation. Notice that in the quantum
vacuum states the dual photon becomes massive which is equivalent to the
confinement of the original abelian electric photons.

\startappendix

\Appendix{Some Lie algebra conventions}

In this Appendix we give a brief review of particular details of Lie
algebras that we will need. For more
details on Lie algebras the reader may consult Refs.\cite{alg}.

Let $ \{ H^i \} $ be a maximal set of simultaneously diagonalizable,
mutually commuting generators, $[H^i, H^j] =0$. The indices $i, j$ run
from 1 to $r$, the \textit{rank} of the Lie algebra.
We normalize the Cartan generators to one,
\EQ{\tr (H^i H^j) = \delta^{ij} \ ,
\label{nncg}}
and often think of the $r$-vector $\boldsymbol H$.
The remainder of the
generators are the step operators $E_{\boldsymbol\alpha}$ with
\EQ{
[\boldsymbol H, E_{\boldsymbol\alpha}] = \boldsymbol\alpha E_{\boldsymbol\alpha} \ .
}
The normalization condition \eqref{nncg} makes the length squared
of any long root to be equal to $2$.

We will denote a set of simple roots as $\boldsymbol\alpha_j$, $
j=1,\ldots,r$. These span the root lattice $\Lambda_R$. The lowest
root is then denoted as $\boldsymbol\alpha_0$. The co-roots are defined via
\EQ{
\boldsymbol\alpha^* \equiv \frac{2}{\boldsymbol\alpha^2}\boldsymbol \alpha \ .
}
and these span the co-root lattice $\Lambda_R^*$. The weight lattice $\Lambda_W$
is dual to the co-root lattice and is spanned by the fundamental
weights $\boldsymbol\omega_j$ where
\EQ{
\boldsymbol\omega_i\cdot\boldsymbol\alpha_j^*=\delta_{ij}\ .
}
Similarly one can define the co-weight lattice $\Lambda_W^*$ which is
dual to the root lattice and is spanned by the co-weights $\boldsymbol\omega_i^*$
where
\EQ{
\boldsymbol\omega_i^* \equiv \frac{2}{\boldsymbol\alpha_i^2}\boldsymbol\omega_i \ .
}

We will also need to define the dual Kac labels, or co-marks, $k_i^*$. By definition
$k_0^*=1$ and the remaining co-marks are given by the expansion of the
lowest co-root in terms of the co-simple-roots:
\EQ{
\boldsymbol\alpha_0^*=-\sum_{i=1}^rk_i^*\boldsymbol\alpha_i^*\ .
}
Finally
\EQ{
c_2 \equiv \sum_{i=0}^{r} k_i^*
}
is the \textit{dual Coxeter number}. (The Kac labels, or marks, and Coxeter number are
similarly defined but will not be needed here.)

In Table 3 we summarize all the Lie algebra data that we
need. As well as listing the dual Kac labels and dual Coxeter number we also
list the root lengths $\boldsymbol\alpha_j^2$ for $j=0,\ldots,r$. (Note that the
set of dual Kac labels and root lengths are ordered in the same way.)

\begin{table}
\[
\begin{array}{ccccc} \hline\hline
\Rsm G & g & 
\Rsm c_2 & \{k_i^*\} & \{\boldsymbol\alpha_i^2\} \\ \hline
\Rsm \SU(r+1) & a_r & r+1 & \{1,1,\ldots,1\} & \{2,\ldots,2\}\\
\Rsm \SO(2r+1)\qquad& b_r\qquad & 2r-1\qquad & \{1,1,2,\ldots,2,1\}\qquad &\{2,\ldots,2,1\}\qquad\\
\Rsm \USp(2r)& c_r &  r+1 &\{1,\ldots,1\} &\{2,1,\ldots,1,2\}\\
\Rsm \SO(2r)& d_r &  2r-2 & \{1,1,2,\ldots,2,1,1\}& \{2,\ldots,2\}\\
\Rsm \G_2 &g_2 & 4 &\{1,2,1\} &\{2,2,2/3\}\\
\Rsm \F_4 &f_4 & 9 &\{1,2,3,2,1\} &\{2,2,2,1,1\}\\
\Rsm \E_6 &e_6 & 12 &\{1,1,1,2,2,2,3\}& \{2,\ldots,2\}\\
\Rsm \E_7 & e_7 & 18 &\{1,1,2,2,2,3,3,4\}& \{2,\ldots,2\}\\
\Rsm \E_8 & e_8 & 30 &\{1,2,2,3,3,4,4,5,6\}& \{2,\ldots,2\}\\
\hline\hline
\end{array}
\]
\caption{\small Lie algebra data.}
\end{table}

\newpage

\Appendix{The monopole collective coordinate measure}\label{sec:B}

In this appendix we briefly discuss the measure
for integrating over the collective coordinates of a fundamental monopole.
A fundamental monopole has a moduli space that is identical to the BPS
monopole in $\SU(2)$. Therefore, it is parametrized by $a_\mu$, the
position in ${\mathbb R}^3$ and by the $\U(1)$ phase angle
$0\leq\Omega\leq2\pi$. Along with this, there are two Grassmann collective
coordinates $\xi_\alpha$, corresponding to the two supersymmetric
zero modes. The measure for integrating over the monopole moduli space
is obtained in the standard way by changing variables
in the path integral from field fluctuations around the monopole
to the monopole's collective coordinates:
\EQ{
\int d\mu_{\rm mon} \ = \
\mu^3 e^{-S} \ \int \frac{d^3 a}{(2\pi)^{\frac{3}{2}}} J_a
 \int_{0}^{2\pi} \frac{d \Omega}{(2\pi)^{\frac{1}{2}}}
J_{\Omega} \int \frac{d^2 \xi}{J_F}
\ , \label{bmpr}
}
where $S$ is the monopole action \eqref{monact} and
$\mu$ is the Pauli-Villars mass scale.
The Jacobian factors $J_a$ and $J_F$ were calculated in \cite{DKMTV}:
\EQ{
J_a= ({\rm Re} S)^{\frac{3}{2}} \ , \qquad
J_F=2{\rm Re} S \ ,
\label{bjxjf}}
and $S$ is the monopole action.
The remaining Jacobian $J_{\Omega}$ is given by
\EQ{
J_{\Omega} = \frac{2\pi R ({\rm Re}S)^{\frac{1}{2}}}{\boldsymbol\alpha \cdot \langle\boldsymbol\varphi\rangle
} \ .
\label{bjom}
}
To derive this, we start with the general expression for the bosonic
zero mode
\EQ{
Z_m = \frac{\partial v_m^{\sst(\Omega)}}{\partial \Omega} + D_m \Lambda \ ,
}
where $v_m^{(\Omega)}$ is the $\Omega$-rotated monopole solution in the
singular gauge,
\EQ{
v_m^{\sst(\Omega)} = e^{i \Omega \tau^3} v_m e^{-i \Omega \tau^3} \ ,
}
and $D_m \Lambda$ is added to keep the zero mode in the covariant
background gauge. Since
\EQ{
\frac{\partial v_m^{\sst(\Omega)}}
{\partial \Omega} \ = \ i \left[\frac{1}{2} \boldsymbol\alpha^*
\cdot \boldsymbol H \ , \ v_m \right] \ ,
}
the choice of $\Lambda$ is obvious (recall \eqref{bhiggsvev}):
\EQ{
\Lambda \ = \ \frac{2 \pi R}{\boldsymbol\alpha \cdot \langle\boldsymbol\varphi\rangle}
\ \Phi^c \tau^c \ - \ \frac{1}{2} \boldsymbol\alpha^* \cdot \boldsymbol H \ .
}
This gives
\EQ{
Z_m \ = \ \frac{2\pi R}{\boldsymbol\alpha \cdot \langle\boldsymbol\varphi\rangle}
\ D_m (\Phi^c \tau^c) \ = \
\frac{2\pi R}{\boldsymbol\alpha \cdot\langle\boldsymbol\varphi\rangle} \ v_{m0} \ ,
}
and
\EQ{
J_{\Omega} \ \equiv \ \sqrt{\langle Z_m | Z_m \rangle} \ = \
 \frac{2\pi R ({\rm Re} S)^{\frac{1}{2}}}{\boldsymbol\alpha \cdot \langle\boldsymbol\varphi\rangle
} \ .
}

Gathering all factors together, we find that the measure is
\EQ{
\frac{\mu^3 R}{g^2}\cdot \frac{2}{\boldsymbol\alpha^2} \cdot e^{-S}
\int d^3 a \ d\Omega \ d^2 \xi  \ . \label{bmeas}
}
In contradistinction with the three-dimensional calculation of \cite{DKMTV}, our
present calculation is locally four-dimensional, {\it i.e.\/}~in the path integral
we have integrated over the fluctuations around the monopole configuration
in ${\mathbb R}^3\times S^1$. Thus, the UV-regularized
determinants over non-zero eigenvalues of the
quadratic fluctuation operators cancel between fermions and bosons
due to supersymmetry as in Ref.~\cite{Adda}.
The ultra-violet divergences are regularized in the Pauli-Villars scheme,
which explains the appearance of the Pauli-Villars mass scale
$\mu$.


\begin{thebibliography}{99}

{\small


\bibitem{NSVZtwo}{V.A. Novikov, M.A. Shifman, A.I. Vainshtein and
V.I. Zakharov, Nucl. Phys. {\bf B260} (1985) 157.}

\bibitem{Shifman:1988ia}
M.A.~Shifman and A.I.~Vainshtein,
Nucl. Phys. {\bf B296} (1988) 445.

\bibitem{Morozov}
A. Morozov, M. Ol'shanetsky
and M. Shifman, Nucl. Phys. {\bf B304} (1988) 291.

\bibitem{FP}
D.~Finnell and P.~Pouliot,
Nucl. Phys. {\bf B453} (1995) 225
{\tt [hep-th/9503115]}.

\bibitem{Aharony:2000ti}
O.~Aharony, S.~S.~Gubser, J.~Maldacena, H.~Ooguri and Y.~Oz,
%``Large N field theories, string theory and gravity,''
Phys.\ Rept.\  {\bf 323} (2000) 183
{\tt[hep-th/9905111]}.
%%CITATION = HEP-TH 9905111;%%
%\href{\wwwspires?eprint=HEP-TH/9905111}{SPIRES}

\bibitem{Intriligator:1996au}
K.~Intriligator and N.~Seiberg,
%``Lectures on supersymmetric gauge theories and electric-magnetic  duality,''
Nucl.\ Phys.\ Proc.\ Suppl.\  {\bf 45BC} (1996) 1
{\tt [hep-th/9509066]}.
%%CITATION = HEP-TH 9509066;%%
%\href{\wwwspires?eprint=HEP-TH/9509066}{SPIRES}

\bibitem{DK}
N.M.~Davies and V.V.~Khoze, JHEP {\bf 0001} (2000) 015
{\tt [hep-th/9911112]}.

\bibitem{NSVZone}{V.A. Novikov, M.A. Shifman, A.I. Vainshtein and
V.I. Zakharov, Nucl. Phys. {\bf B229} (1983) 394;
Nucl. Phys. {\bf B229} (1983) 407.}

\bibitem{ARV}{G.C. Rossi and G. Veneziano, Phys. Lett. {\bf 138B} 195;
D. Amati, G.C. Rossi and G. Veneziano,
Nucl. Phys. {\bf B249} (1985) 1.}

\bibitem{Amati}
D.~Amati, K. Konishi, Y. Meurice, G.C. Rossi and G. Veneziano,
Phys. Rept. {\bf 162} (1988) 169.

\bibitem{ADS}
I.~Affleck, M.~Dine and N.~Seiberg,
Nucl. Phys. {\bf B241} (1984) 493.

\bibitem{FS}
J.~Fuchs and M.G.~Schmidt,
Z. Phys. {\bf C30} (1986) 161.

\bibitem{KS}{A. Kovner and M. Shifman, Phys. Rev {\bf D 56} (1997) 2396
{\tt [hep-th/9702174]}.}


\bibitem{SVrev}
M.~Shifman and A.~Vainshtein,
{\tt hep-th/9902018}.


\bibitem{HKLM}
T.~J.~Hollowood, V.~V.~Khoze, W.~Lee and M.~P.~Mattis,
%``Breakdown of cluster decomposition in instanton calculations of the  gluino condensate,''
Nucl.\ Phys.\  {\bf B570} (2000) 241
{\tt[hep-th/9904116]}.
%%CITATION = HEP-TH 9904116;%%
%\href{\wwwspires?eprint=HEP-TH/9904116}{SPIRES}

\bibitem{MO}
N. Dorey, V.V. Khoze and M.P. Mattis,
Phys.~Rev.~{\bf D54} (1996) 2921 {\tt [hep-th/9603136]};
Phys.~Rev.~{\bf D54} (1996) 7832 {\tt [hep-th/9607202]}.

\bibitem{DHKMV}
N. Dorey, T.J. Hollowood, V.V. Khoze, M.P. Mattis and
S. Vandoren,
Nucl. Phys. {\bf B552} (1999) 88
{\tt [hep-th/9901128]}.

\bibitem{DHKM}
N.M.~Davies, T.J.~Hollowood, V.V.~Khoze and M.P.~Mattis,
Nucl. Phys. {\bf B559} (1999) 123
{\tt [hep-th/9905015]}.

\bibitem{Belavin:1979fb}
A.A.~Belavin, V.A.~Fateev, A.S.~Schwarz and Y.S.~Tyupkin,
Phys. Lett. {\bf 83B} (1979) 317.

\bibitem{Fateev:1979dc}
V.A.~Fateev, I.V.~Frolov and A.S.~Shvarts,
Nucl.\ Phys.\ {\bf B154} (1979) 1.

\bibitem{Berg:1979uq}
B.~Berg and M.~Luscher,
Commun.\ Math.\ Phys.\ {\bf 69} (1979) 57.

\bibitem{Callan:1978gz}
C.G.~Callan, R.~Dashen and D.J.~Gross,
Phys.\ Rev.\ {\bf D17} 2717 (1978) 2717.

\bibitem{Osborn}{H. Osborn, Ann. Phys. {\bf 135} (1981) 373.}

\bibitem{LY}{K. Lee and P. Yi,
Phys. Rev. {\bf D56} (1997)
3711 {\tt [hep-th/9702107]}.}

\bibitem{KL}{K. Lee,
Phys. Lett. {\bf B426} (1988) 323
    {\tt [hep-th/9802012]}.}

\bibitem{Nahmtwo}{W. Nahm, {\sl Self-dual Monopoles and Calorons\/},
in: Lecture Notes in Physics 201, Eds. G. Denado {\it et. al.} 1984,
p. 189.}

\bibitem{Garland}{H. Garland and M.K. Murray, Comm. Math. Phys. {\bf 120}
(1988) 335}.


\bibitem{LL}{
 K. Lee and C. Lu,
 Phys. Rev. {\bf D58} (1988) 025011
    {\tt [hep-th/9802108]}.}

\bibitem{KvB}{T.C. Kraan and P. van Baal,
 Phys. Lett. {\bf B428} (1998) 268 {\tt [hep-th/9802049]};
 Nucl. Phys. {\bf B533} (1998) 627 {\tt [hep-th/9805168]};
Phys. Lett. {\bf B435} (1998) 389  {\tt [hep-th/9806034]}.}


\bibitem{HS}{B.J. Harrington and H.K. Shepard, Phys. Rev. {\bf D17} (1978) 2122.}

\bibitem{Rossi:1979qe}
P.~Rossi,
Nucl.\ Phys.\ {\bf B149} (1979) 170.

\bibitem{GPY}{D.J. Gross, R.D. Pisarski and L.G. Yaffe,
Rev. Mod. Phys. {\bf 53} (1981) 43.}


\bibitem{SWthree}
N.~Seiberg and E.~Witten,
{\tt hep-th/9607163}.


\bibitem{Aharony:1997bx}
O.~Aharony, A.~Hanany, K.~Intriligator, N.~Seiberg and M.~J.~Strassler,
%``Aspects of N = 2 supersymmetric gauge theories in three dimensions,''
Nucl.\ Phys.\  {\bf B499} (1997) 67
{\tt [hep-th/9703110]}.
%%CITATION = HEP-TH 9703110;%%
%\href{\wwwspires?eprint=HEP-TH/9703110}{SPIRES}


\bibitem{Dorey}
N.~Dorey,  JHEP {\bf 9907} (1999) 021,
{\tt [hep-th/9906011]}.


\bibitem{KAC}V.G.~Kac, {\sl Infinite Lie algebras}, Progress in
Mathematics. Vol 44. Boston: Birkh\"auser 1983.


\bibitem{Witten:1996bn}
E.~Witten,
Nucl.\ Phys.\ {\bf B474} (1996) 343
{\tt [hep-th/9604030]}.


\bibitem{Katz:1997th}
S.~Katz and C.~Vafa,
Nucl.\ Phys.\ {\bf B497} (1997) 196
{\tt [hep-th/9611090]}.
%%CITATION = NUPHA,B497,196;%%

\bibitem{Vafa:1998vs}
C.~Vafa,
Adv.\ Theor.\ Math.\ Phys.\ {\bf 2} (1998) 497
{\tt [hep-th/9801139]}.
%%CITATION = 00203,2,497;%%

\bibitem{Windx}E. Witten, Nucl. Phys. {\bf B202} (1982) 253;
JHEP {\bf 9802} (1998) 006 {\tt [hep-th/9712028]};
A. Keurentjes, A. Rosly and A.V. Smilga,
Phys. Rev. {\bf D58} (1998) 081701 {\tt [hep-th/9805183]}.

\bibitem{Dirac} P.A.M. Dirac, Proc. Roy. Soc. {\bf A133} (1931) 60;
F. Englert and P. Windey, Phys. Rev. {\bf D14} (1976) 2728;
P. Goddard, J. Nuyts and D. Olive, Nucl. Phys. {\bf B125} (1977) 1.

\bibitem{Weinberg:1982ev}
E.J.~Weinberg,
Nucl.\ Phys.\ {\bf B167} (1980) 500;
Nucl.\ Phys.\ {\bf B203} 445 (1982) 445.

\bibitem{CD} S.F. Cordes and M. Dine, Nucl. Phys. {\bf B273} (1986) 581.

\bibitem{thm}{G. 't Hooft, Nucl. Phys. {\bf B79} (1974) 276.}

\bibitem{polm}{A.M. Polyakov, JETP Lett. {\bf 20} (1974) 194.}

\bibitem{Bog}{ E.B. Bogomol'nyi, Sov. J. Phys {\bf 24} (1977) 97.}

\bibitem{PS}{ M.K. Prasad and C.M. Sommerfield, Phys. Rev. Lett.
{\bf 35} (1975) 760.}


\bibitem{Seiberg93}
{N. Seiberg, Phys. Lett. {\bf 318B} (1993) 469 {\tt [hep-th/9309335]}.}

\bibitem{DKMTV}{N. Dorey, V.V. Khoze, M.P. Mattis, D. Tong and
S. Vandoren,  Nucl. Phys. {\bf B502} (1997) 94
{\tt [hep-th/9703228]}.}


\bibitem{Gomez:1997mq}
C.~Gomez,
{\tt hep-th/9706131}.
%%CITATION = HEP-TH 9706131;%%

\bibitem{Gomez:1997}
C.~Gomez and R.~Hernandez,
Nucl.\ Phys.\ Proc.\ Suppl.\ {\bf 61A} (1998) 172
{\tt [hep-th/9703036]};
Int.\ J.\ Mod.\ Phys.\ {\bf A12} (1997) 5141
{\tt [hep-th/9701150]}.
%%CITATION = IMPAE,A12,5141;%%
\bibitem{MW}
E.~Martinec and N.~Warner,
%``Integrable systems and supersymmetric gauge theory,''
Nucl.\ Phys.\  {\bf B459} (1996) 97
{\tt [hep-th/9509161]}.
%%CITATION = HEP-TH 9509161;%%
%\href{\wwwspires?eprint=HEP-TH/9509161}{SPIRES}


\bibitem{TJH}
T.~J.~Hollowood,
%``Strong coupling N = 2 gauge theory with arbitrary gauge group,''
Adv.\ Theor.\ Math.\ Phys.\  {\bf 2} (1998) 335
{\tt [hep-th/9710073]}.
%%CITATION = HEP-TH 9710073;%%
%\href{\wwwspires?eprint=HEP-TH/9710073}{SPIRES}



\bibitem{AHW}
I.~Affleck, J.A.~Harvey and E.~Witten,
Nucl.\ Phys.\ {\bf B206} (1982) 413.
%%CITATION = NUPHA,B206,413;%%

\bibitem{alg}
R. Gilmore, {\sl Lie groups, Lie algebras and some of their
representations}, Wiley, 1974;
H. Samelson, {\sl Notes on Lie algebras}, Springer 1990;
R. N. Cahn, {\sl Semisimple Lie algebras and their representations},
J. Adams, {\sl Lectures on exceptional
Lie groups}, Chicago, 1996, ed. Z. Mahmud, M. Mimura
Benjamin/Cummings, 1984.

\bibitem{Adda}{ A. D'Adda and P. Di Vecchia,
Phys. Lett. {\bf 73B} (1978) 162.}

}

\end{thebibliography}
\end{document}